\documentclass[twocolumn,english,aps,PRB,groupedaddress,floats]{revtex4}
\usepackage[T1]{fontenc}
\usepackage[latin9]{inputenc}
\usepackage{graphicx}



\usepackage{babel}

\begin{document}

\title{Electronic structure of Fe$_{1.04}$(Te$_{0.66}$Se$_{0.34}$)}

\author{Fei Chen$^{1}$, Bo Zhou$^{1}$, Yan Zhang$^{1}$, Jia Wei$^{1}$, Hong-Wei Ou$^{1}$, Jia-Feng Zhao$^{1}$, Cheng He$^{1}$, Qing-Qin
Ge$^{1}$, Masashi Arita$^{2}$, Kenya Shimada$^{2}$, Hirofumi
Namatame$^{2}$, Masaki Taniguchi$^{2}$, Zhong-Yi Lu$^{3}$,  Jiangping Hu$^{4}$, Xiao-Yu Cui$^{5}$, D. L. Feng$^{1}$}

\email{dlfeng@fudan.edu.cn}

\affiliation{$^{1}$Department of Physics, Surface Physics Laboratory (National
Key Laboratory), and Advanced Materials Laboratory, Fudan University,
Shanghai 200433, P. R. China}

\affiliation{$^{2}$Hiroshima Synchrotron Radiation Center and Graduate School
of Science, Hiroshima University, Hiroshima 739-8526, Japan.}

\affiliation{$^3$Department of Physics, Renmin University of China,
Beijing 100872, P. R.  China}

\affiliation{$^4$ Department of Physics, Purdue University, West
Lafayette, IN 47907, USA}

\affiliation{$^5$ Swiss Light Source, Paul-Scherrer Institut, 5232
Villigen, Switzerland}

\date{\today}
\begin{abstract}
We report the electronic structure of the iron-chalcogenide
superconductor, Fe$_{1.04}$(Te$_{0.66}$Se$_{0.34}$), obtained with high resolution angle-resolved photoemission spectroscopy and density functional calculations. In photoemission measurements, various photon energies and polarizations are exploited to study the Fermi surface topology and symmetry properties of the bands. The measured band structure and their symmetry characters qualitatively agree with our density function theory calculations of Fe(Te$_{0.66}$Se$_{0.34}$), although the band structure is renormalized by about a factor of three. We find that the electronic structures of this iron-chalcogenides and the iron-pnictides have many aspects in common, however, significant differences exist near the $\Gamma$-point. For
Fe$_{1.04}$(Te$_{0.66}$Se$_{0.34}$), there are clearly separated three bands with distinct even or odd symmetry that cross the Fermi energy ($E_F$) near the zone center, which contribute to three hole-like Fermi surfaces. Especially, both experiments and calculations show a hole-like elliptical Fermi surface at the zone center. Moreover, no sign of spin density wave was observed in the electronic structure and susceptibility measurements of this compound.
\end{abstract}

\pacs{74.25.Jb,74.70.-b,79.60.-i,71.20.-b}

\maketitle

\section{Introduction}

The discovery of superconductivity with the superconducting
transition temperature ($T_{c}$) up to 55 K in iron-pnictides
\cite{Hosono,XHChen,ZXZhao} has generated great interests. The FeAs
layer is considered as the key structure for superconductivity in
systems ranging from SmO$_{1-x}$F$_x$FeAs,
Ba$_{1-x}$K$_{x}$Fe$_{2}$As$_{2}$ \cite{Johrendt1,Johrendt2}, to
LiFeAs \cite{Clarke,Tapp}. Recently, certain iron-chalcogenides,
\textit{eg.} Fe$_{1+x}$Se, Fe$_{1+y}$Te$_{1-x}$Se$_{x}$
\cite{Wu1,Wu2}, have been found to be superconducting as well.
Fe$_{1+x}$Se shows superconductivity at 8 K under ambient pressure
\cite{Wu1} and 37 K under a 7 GPa hydrostatic pressure
\cite{Prassides}, which is comparable to
Ba$_{1-x}$K$_{x}$Fe$_{2}$As$_{2}$ (T$_{c}$=38 K) \cite{Johrendt2}.
Because iron-chalcogenides  do not involve arsenic, it would be
particularly important for applications. Furthermore, although Fe
$3d$ orbitals play a vital role in the iron-based high temperature
superconductors,  the anions seem also play an important role on
various aspects,  noting LaOFeP possesses a T$_{c}$ of merely 5 K.
Besides the size effect, the polarizability of the anions has even
been suggested to be crucial for the superconductivity
\cite{Sawatzky}. Therefore, iron-chalcogenides provide an
opportunity to study the role of anions in iron-based
superconductors.

\begin{figure}[t!]
\centering \includegraphics[width=6cm]{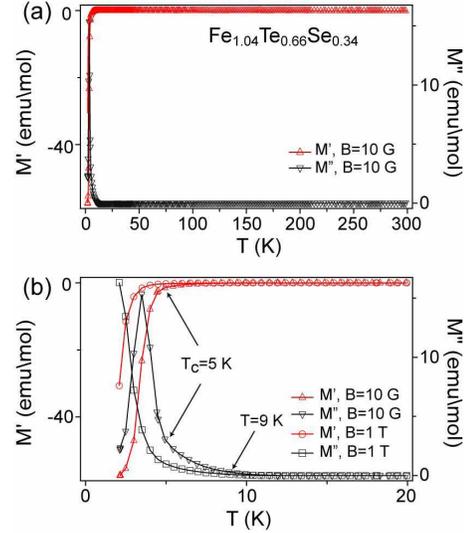} \caption{(color
online) (a)  Real and imaginary parts of the magnetic susceptibility
of Fe$_{1.04}$(Te$_{0.66}$Se$_{0.34}$) single crystal at 10 Gauss.
The data were taken with the zero field cool (ZFC) procedure. (b)
The low temperature magnetic susceptibility data  at 10 Gauss and 1
T magnetic field.} \label{Setup}
\end{figure}

The iron-pnictides and iron-chalcogenides have many things in
common. The FeSe(Te) layer in Fe$_{1+y}$Te$_{1-x}$Se$_{x}$ is isostructural to the FeAs or FeP layer in iron-pnictides. Moreover, the phase diagram of Fe$_{1+y}$Te$_{1-x}$Se$_{x}$ resembles that of the iron-pnictides, where the competition between magnetism and superconductivity has been
observed in both cases. The undoped Fe$_{1+y}$Te exhibits a spin density wave (SDW) ground state. With sufficient
selenium doping, the SDW is suppressed, and the
superconductivity occurs at a T$_{c}$ as high as 15 K \cite{Wu2}.

\begin{figure*}[t!]
\includegraphics[width=17cm]{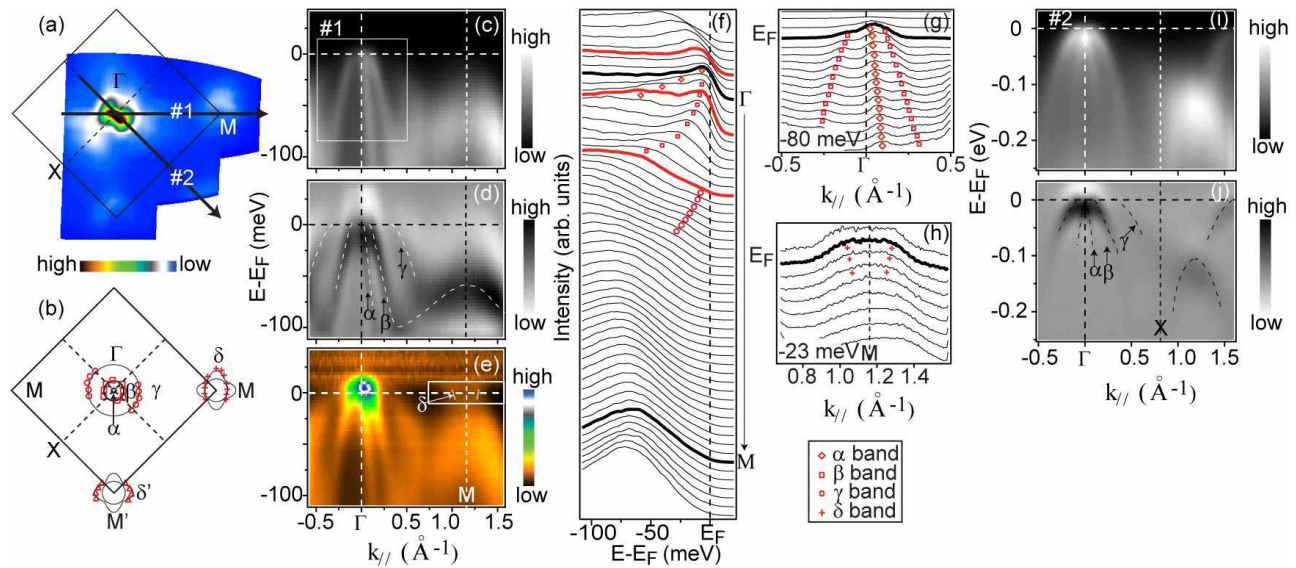}
\caption{(color online) (a) Photoemission intensity distribution
integrated over $[E_F-5~meV, E_F+5~meV]$ window for
Fe$_{1.04}$(Te$_{0.66}$Se$_{0.34}$). (b) The Fermi surfaces are
constructed based on the measured Fermi crossings, which are labeled
by squares, circles, crosses and triangles  for the $\beta$,
$\gamma$, $\delta$ and $\delta^{'}$ bands respectively. (c) The
photoemission intensity along the cut \#1 in the $\Gamma-$M
direction, and (d) its second derivative with respect to energy. (e)
The data in panel c is repotted after dividing the angle integrated
energy distribution curve. (f) The EDC's for data in panel c.  (g)
The MDC's in the box area of panel c. (h) The MDC's in the box area of panel e. The crosses mark  the feature positions  based on a double Lorenzian fit.(i) The photoemission intensity along the cut \#2 in the
$\Gamma-$X direction, and (j) its second derivative. Data were taken
with circularly polarized 22~eV photons at HSRC.} \label{FS}
\end{figure*}

There are also critical differences between the iron-pnictides and
iron-chalcogenides, in particular, between their structures of
magnetic ordering. A common collinear commensurate antiferromagnetic
(AFM)  ordering has been identified in all iron-pnictides. However,
the magnetic state of the Fe$_{1+y}$Te$_{1-x}$Se$_{x}$ family  has
a bi-collinear commensurate or incommensurate antiferromagnetic
ordering depending on the concentration of the interstitial iron
\cite{Mao,PCDai}. It is still in a heated debate about the origin of
magnetic ordering in iron-based superconductors. While models based
on local moments have been suggested to understand both magnetic
orderings \cite{Ma2008,Fang2008,Dai2009, ma2009, chen2009, cenke},
the collinear AFM in the iron-pnictides in principle can originate
from nesting mechanism between the hole pockets at $\Gamma$ and  the
electron pockets at $M$ \cite{Dong}, but the bi-collinear magnetic
structure is inconsistent with this picture since there is no Fermi
surface at $X$. Is there a connection between the electronic
structure and magnetic ordering in the iron-chalcogenides? If there
is, what is the connection?  The answers of these fundamental
questions  require a deep understanding of the electronic structures of iron-chalcogenides. However, there is  just few  data reported on the electronic structure of iron-chalcogenides \cite{Takahashi2009}.

In this Article, we
investigate the electronic structure of
Fe$_{1.04}$Te$_{0.66}$Se$_{0.34}$ with high resolution
angle-resolved photoemission spectroscopy (ARPES) and band
calculation. The  measured Fermi surfaces and the band structure are
identified and found to qualitatively agree with the density
function theory (DFT) calculations. The orbital characters of
individual bands are studied by polarization-dependence studies and
found to agree with the calculation as well. No obvious effect of
the fluctuating SDW is observed on the electronic structure.
Furthermore, we found that although most aspects of the electronic
structure of this iron-chalcogenide are similar to the
iron-pnictides, there are clearly three separated  bands  at the
zone center for the iron-chalcogenides, while there appear just two
separated features for the normal state of iron-pnictides.  Moreover, the symmetry properties of the iron-chalcogenide bands near the zone center are
different from those of the iron-pnictides. The difference and
similarity between the iron-pnictides and iron-chalcogenides in
their electronic structure may shed light on our understanding of
the role of anions and the superconductivity in iron-based
superconductors.

\section{Material and experimental setup}

Fe$_{1.04}$(Te$_{0.66}$Se$_{0.34}$) single crystal was synthesized
with the NaCl/KCl-flux method. Fe powder, Te powder and Se powder
were weighed according to the ratio of Fe:Te:Se=1:0.7:0.3 (mole),
and pressed into thin plates. Then FeTe(Se) polycrystal was acquired
by reacting the plate in an evacuated quartz tube at 1173 K for 24
hours. FeTe(Se) polycrystal and the NaCl/KCl-flux were weighed
according to the ratio of FeTe(Se): NaCl/KCl=1:10 (mass). They were
thoroughly grounded into a mixture, and loaded into an evacuated
quartz tube. The tube was kept at 1223 K for 24 hours and then
slowly cooled to 873 K in 100 hours. Finally the quartz tube was
cooled in the furnace after shutting off the power.
Fe$_{1.04}$(Te$_{0.66}$Se$_{0.34}$) single crystal was obtained
after dissolving the flux in deionized water. The element
compositions of this single crystal were determined through
energy-dispersive x-ray (EDX) analysis with dense sampling spots
across a 0.4$\times$0.4 mm$^{2}$ surface area. The EDX result shows
that the sample is homogeneous, and the maximal deviation of its
compsitions is within 1.8\%. The temperature dependence of the
magnetic susceptibility (Fig.~1) does not show any signs of SDW or
structural transition. The resistivity data indicate that crystal
reaches the zero resistance at about 9 K. However, the
susceptibility measurements show that although regions of the sample
become superconducting at 9 K, it reaches a bulk superconducting
state at 2 K, with a transition width less than 3 K
(10${\%}$-90${\%}$). This indicates that the bulk of the single
crystal is quite homogeneous. With 1 T magnetic field,
superconductivity is suppressed, and there is no sign of
field-induced meta-magnetic transition.

\begin{figure}[t!]
\includegraphics[width=8.5cm]{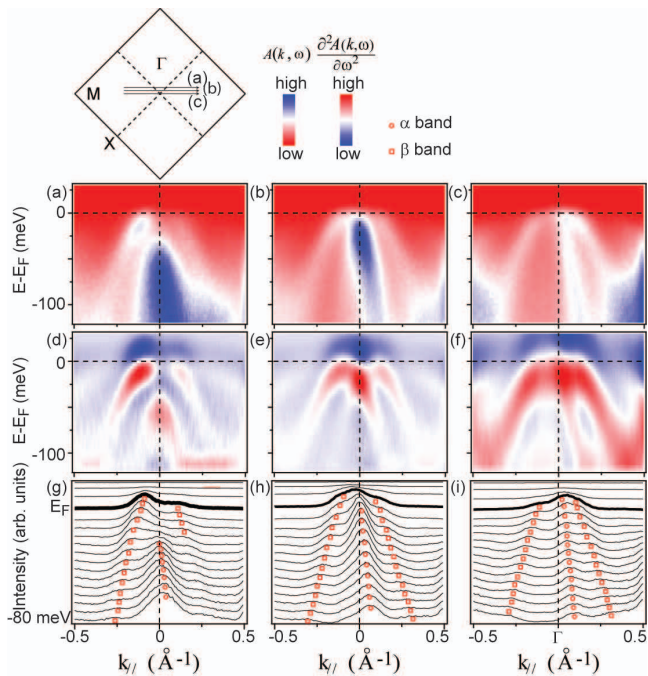}
\caption{(color online) (a, b, c) Photoemission data taken  with 22~eV circularly polarized photons at HSRC along
three momentum cuts as indicated in the Brillouin zone sketch. (d,
e, f) Second derivative with respect to energy for data in panels a,
b, c respectively. (g, h, i) Momentum distribution curves near $E_F$
for data in panels a, b, c respectively.} \label{LDAorbital}
\end{figure}

\begin{figure}[t!]
\includegraphics[width=8.5cm]{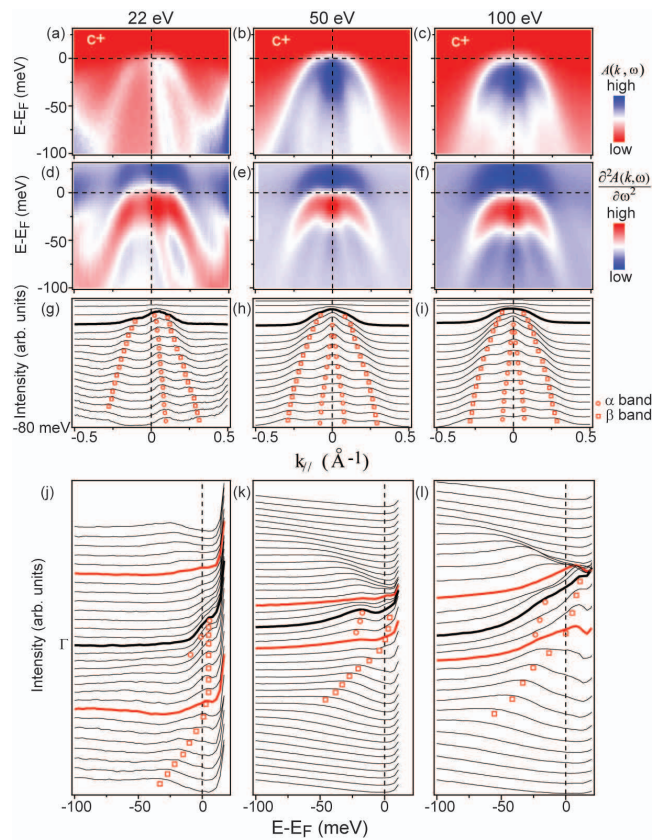}
\caption{(color online) (a, b, c) Photoemission data taken  with
22~eV light at HSRC, and 50~eV, and 100~eV circularly polarized
light at SLS respectively. All three momentum cuts cross the
$\Gamma$-Z line ($k_x$=0, $k_y$=0, $k_z$) in the reciprocal space.
(d, e, f) Second derivative with respect to energy for data in
panels a, b, c respectively. (g, h, i) Momentum distribution curves
near $E_F$ for data in panels a, b, c respectively. (j, k, l) Energy
distribution curves for data in panels a, b, c respectively after
divided by the Fermi-Dirac distribution function.} \label{LDA1}
\end{figure}

The photoemission data have been taken with Scienta R4000 electron
analyzers at Beamline 9 of Hiroshima synchrotron radiation center
(HSRC) and the Surface and Interface Spectroscopy (SIS) Beamline of
Swiss Light Source (SLS). The typical angular resolution is 0.3
degree, and the typical energy resolution is 15 meV. The sample was
cleaved \textit{in situ}, and measured under ultra-high-vacuum
better than $5\times10^{-11}$ \textit{torr}. The sample aging
effects are carefully monitored to ensure they do not cause
artifacts in our analyses and conclusions. The SIS beamline is
equipped with an elliptically polarized undulator (EPU), which could
switch the photon polarization between horizontal, vertical, or
circular mode. This facilitates the polarization dependence studies,
which is useful in determining the orbital characters of the
bands~\cite{Zhang09B}.

\section{Band structure and Fermi surface}

\begin{figure}[t!]
\includegraphics[width=8.5cm]{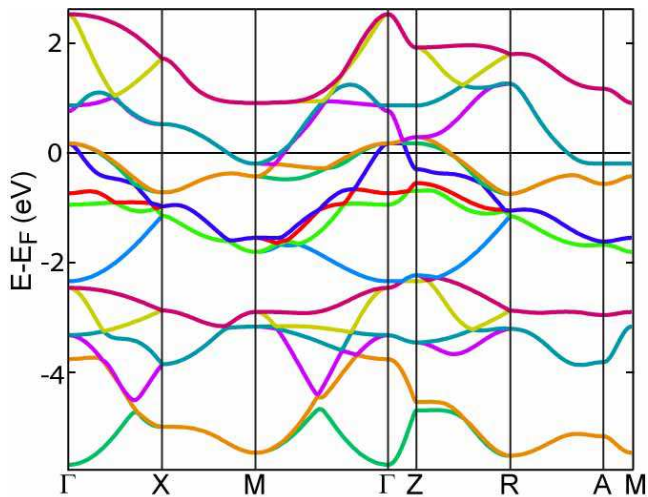}
\caption{(color online) The calculated electronic band structure of
FeTe$_{0.66}$Se$_{0.34}$ along high-symmetry lines in the
irreducible Brillouin zone.} \label{LDAband}
\end{figure}

The photoemission intensities distribution of
Fe$_{1.04}$(Te$_{0.66}$Se$_{0.34}$) at the Fermi energy is shown in
Fig.~2(a). The data were taken at 15 K with 22~eV photons. Similar
to the iron-pnictides, the spectral weight is mostly located around
$\Gamma$  and M. In order to resolve the details of the Fermi
crossings, Fig.~\ref{FS}(c) shows the photoemission intensity along
the cut~\#1 in the $\Gamma-$M direction, several bands could be
resolved. For a better visualization of the bands, Fig.~\ref{FS}(d)
shows the second derivative with respect to energy for the data in
Fig.~\ref{FS}(c). Three bands, $\alpha$, $\beta$, and $\gamma$,
could be clearly identified. The top of the $\alpha$ band is very
close to the Fermi energy, but  it is hard to judge whether it
crosses $E_F$ based on the energy distribution curves (EDC's) in
Fig.~\ref{FS}(f).  By judging from the momentum distribution curves
(MDC's) in Fig.~\ref{FS}(g) near $\Gamma$, one finds that it crosses
the Fermi level with a very small-sized Fermi surface. In order to
check whether the spectral weight around M represents band
crossings, the data in Fig.~\ref{FS}(c) is renormalized by its
angular integrated spectrum and shown in Fig.~\ref{FS}(e). In this
way, another band, $\delta$, is resolved. The MDC's in the boxed
region are shown in Fig.~\ref{FS}(h), where one observes an
electron-pocket type of dispersion. This is similar to the
BaFe$_2$As$_2$ \cite{Yang09}, the $\delta$ band is quite weak in
such an experimental geometry due to the strong orbital-dependence
of the matrix element \cite{Zhang09B}. Similarly, Figs.~\ref{FS}(i)
and (j) show the photoemission intensity and its second derivative
plot along the $\Gamma$-X direction, where $\alpha$, $\beta$, and
$\gamma$ bands are observed. Therefore, there are totally three
bands near the $\Gamma$-point, and  they all cross the Fermi surface
and form three hole pockets. Based on the identified band
dispersions, the Fermi crossings are determined and shown in
Fig.~\ref{FS}(b). We note because of symmetry constraints, only
crossings for one elliptical Fermi surface could be observed around
M or M' \cite{Zhang09B}. The experimental Fermi surfaces are
determined by fitting the Fermi crossings with symmetry in
consideration. Assuming the $\beta$, $\gamma$, and $\delta$ Fermi
surfaces to be cylindrical, one could estimate the electron
concentration based on the Luttinger theorem. We obtained 0.08 holes
per unit cell. This is not inconsistent with the chemical formula,
considering variations of the Fermi surface volume caused by $k_z$
dispersion of the band structure.

\begin{figure}[t!]
\includegraphics[width=8.5cm]{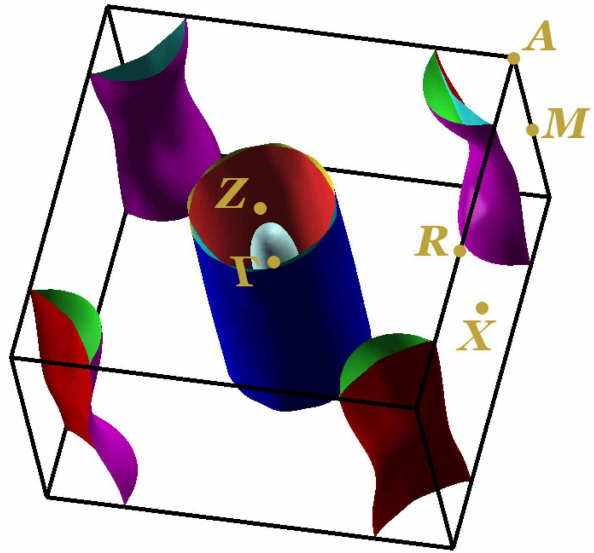}
\caption{(color online) The calculated Fermi surface of
FeTe$_{0.66}$Se$_{0.34}$.} \label{LDAFS}
\end{figure}

To further illustrate the behavior of the $\alpha$ band, Fig.~3
shows three nearby cuts taken with 22~eV photons. When approaching $(k_x=0, k_y=0)$, the $\alpha$ band disperses
rather rapidly with its top in each cut moving towards $E_F$. Based
on the peak positions in the MDC's [Fig.~3(j)], one could observe a
Fermi crossing of  the $\alpha$ band very close to $(k_x=0, k_y=0)$, giving
a small hole-like Fermi surface. However, this Fermi crossing is not
observed at several other photon energies such as 50~eV and 100~eV
(Fig.~4). Since these momentum cuts sample through  $(k_x,
k_y)=(0,0)$ at different $k_z$, the $\alpha$ Fermi surface is thus a
closed pocket. Moreover, in Figs.~4(j-l), the EDC's have been
divided by the temperature-broadened Fermi-Dirac distribution, where
both the $\alpha$ and $\beta$ bands appear to be degenerate within
the experimental resolution at 22~eV. Based on the calculations
below, it suggests that this data cut should be very close to the zone center, $\Gamma$.

\section{Electronic band structure calculation}

To understand the data, we have calculated the electronic band
structure for FeTe$_{0.66}$Se$_{0.34}$. In the calculations the
plane wave basis method was used \cite{pwscf}. We adopted the
generalized gradient approximation of Perdew-Burke-Ernzerhof
\cite{pbe} for the exchange-correlation potentials. The ultrasoft
pseudopotentials \cite{vanderbilt} were used to model the
electron-ion interactions. After the full convergence test, the
kinetic energy cut-off and the charge density cut-off of the plane
wave basis were chosen to be 600~eV and 4800~eV, respectively. The
Gaussian broadening technique was used and a mesh of $16\times
16\times 8$ k-points were sampled for the irreducible Brillouin-zone
integration. The internal atomic coordinates within a cell were
determined by the energy minimization. The doping effect upon
electronic structures was studied by using virtual crystal
calculations.

\begin{figure}[t!]
\includegraphics[width=7.5cm]{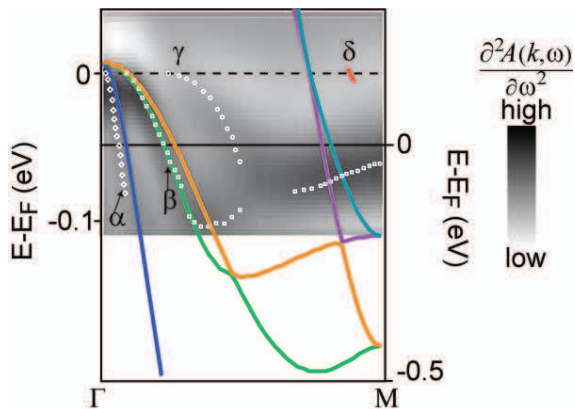}
\caption{(color online) Comparison of the band structure of the
ARPES data and the DFT calculation results along the ${\Gamma}$-M
cut. Note: the energy scale of the calculated band structure at the
right side is 3.125 times of the energy scale of the experimental
data at the left side.} \label{quan}
\end{figure}


Fig.~5 displays the calculated band structure, there are indeed
three bands  near $\Gamma$, and two bands near M that cross the
Fermi energy. In particular, the inner-most band near $\Gamma$ does
show significant dispersion along the $\Gamma-Z$ direction. As a result, our calculations give five Fermi surfaces as shown in Fig.~6. The calculated band structure to a large extent resembles those of the  iron-arsenide superconductors.

Qualitatively, the calculated Fermi surfaces  agree  well with our
experiments. However, there are some important quantitative
discrepancies. Fig.~\ref{quan} illustrates the measured band
structure along $\Gamma$-M as reproduced from Fig.~\ref{FS}(c),
together with the calculated bands. One finds that the size of the
calculated $\gamma$ Fermi surface is much smaller than the measured
one. However, the calculated $\alpha$, $\beta$, and $\gamma$ bands
match the data after scaled by 3.125 and shifted down by 45~meV,
except the Fermi crossings of the measured and calculated $\gamma$
band are different. The scaling factor illustrates the correlation
effects in this material. The experimental Fermi velocity of
$\alpha$, $\beta$, and  $\gamma$ bands are 0.62~$eV$\AA,
0.4~$eV$\AA, and 0.137~$eV$\AA~~respectively. On the other hand, the
measured Fermi surface around M is much smaller than the calculated
ones. Similar to the iron-pnictides \cite{Lu08}, the renormalization
factors of the bands  vary in different regions of the Brillouin
zone.

We note that in order to obtain accurate band renormalization
factors, it is crucial to compare the data with the calculation
conducted \emph{for the same Se doping}. We have calculated the band
structures of FeTe$_{1-x}$Se$_{x}$ with a series of doping, and
found that the band structure around $\Gamma$ evolves rapidly with
increasing Se concentration. Similar conclusion can be drawn from
the published FeSe and FeTe band structures by Subedi and coworkers
\cite{Subedi}.

\section{Polarization dependence}

\begin{figure}[t!]
\includegraphics[width=8.5cm]{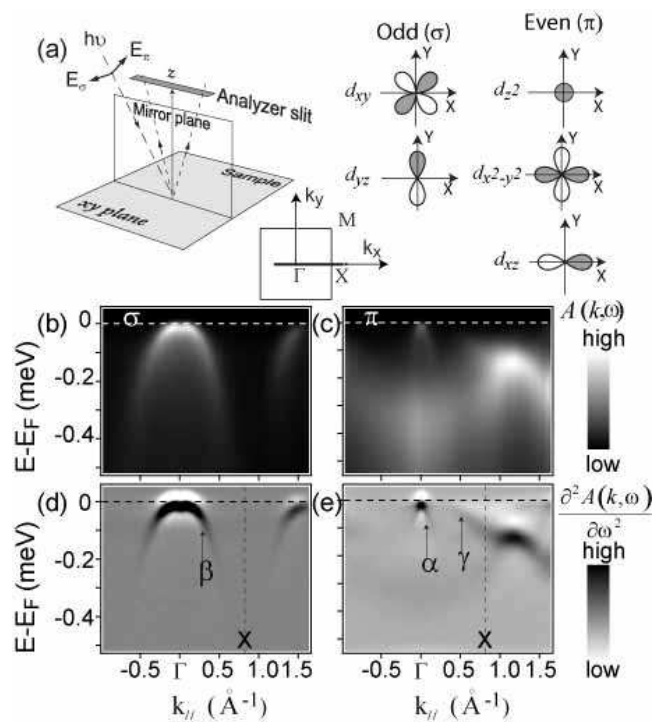}

\caption{(color online) (a) Cartoon of the polarization dependence
experiment, and the symmetry of the orbitals with respect to the
mirror plane defined by sample normal and $\Gamma$-X. (b)  The
photoemission intensity along the $\Gamma$-X direction measured at
the ${\sigma}$ geometry, and (d)  its second derivative with respect
to energy. (c) The photoemission intensity along the $\Gamma$-X
direction measured at the ${\pi}$ geometry, and (e) its second
derivative. Data were taken with 100~eV photons  at SLS, and the
temperature was 10 K.} \label{Pdep}
\end{figure}

For a multi-band and multi-orbital superconductor, it is crucial to
understand the orbital characters of the band structure near  $E_F$.
In photoemission, such information can be obtained to a large extent
in the polarization dependence. Fig.~\ref{Pdep}(a) illustrates two
types of experimental setup with linearly polarized light. The
incident beam and the sample surface normal define a mirror plane.
For the $\pi$ (or $\sigma$) experimental geometry, the electric
field direction ($\bf{\hat{\varepsilon}}$) of the incident photons is in (or out of) the mirror
plane.  The  matrix element of the photoemission process can be described by
$${|M_{f,i}^{\bf{k}}|\propto{\rm{|}}\langle \phi _f^{\bf{k}}
|\bf{\hat{\varepsilon}}\cdot{\bf{r}}|\phi _i^{\bf{k}} \rangle
|^2}$$, where ${\phi _i^{\bf{k}}}$ and ${\phi _f^{\bf{k}}}$  are the initial and final state wavefunctions respectively \cite{ZXShenRev}. In our experimental setup, the momentum of the final-state photoelectron is in the mirror plane, and ${\phi _f^{\bf{k}}}$ can be approximated by a plane wave. Therefore, ${\phi _f^{\bf{k}}}$ is always even with respect to the mirror plane. In the $\pi$   geometry, $(\bf{\hat{\varepsilon}}\cdot{\bf{r}})$ is even, to give a finite photoemission matrix element, $|\phi _i^{\bf{k}} \rangle$ must be even  with respect to the mirror
plane. Thus only even state is probed in the $\pi$ experimental geometry. On the other hand, one could similarly deduce that only odd state is observed in the $\sigma$ geometry.

In contrast to the data measured with circularly polarized light  in
Figs.~\ref{FS}(i) and (j), only the $\beta$ band is observed in the
$\sigma$ geometry  [Figs.~\ref{Pdep}(b) and (d)], while just the
$\alpha$ and $\gamma$ bands are observed in the $\pi$ geometry
[Figs.~\ref{Pdep}(c) and (e)]. Based on the symmetry of different
orbitals illustrated in Fig.~\ref{Pdep}(a), the  $\beta$ band is odd
with respect to the mirror plane, while the $\alpha$ and $\gamma$
bands are even along the $\Gamma$-X direction. Therefore, the
$\beta$ band has to be made of $d_{xy}$ and/or $d_{yz}$ orbitals,
while the $\alpha$ and $\gamma$ bands may be consisted of
$d_{x^2-y^2}$, $d_{z^{2}}$, and/or $d_{xz}$.

These experimental findings of the symmetry properties of the band
structure are well captured by the band structure calculation. In
Fig.~\ref{LDAorbital}, the orbital characters of the bands are shown
by the false color plot. Near the Fermi energy, the $\alpha$ band is
mainly consisted of $d_{xz}$/$d_{yz}$ orbitals, which should be
purely $d_{xz}$  along the $\Gamma-X$ direction. The  $\beta$ band
is consisted of mainly $d_{xz}$ and $d_{yz}$ orbitals, and some
$d_{xy}$ orbitals; while the $\gamma$ band is consisted of
$d_{x^2-y^2}$ orbital. Along the $\Gamma-Z$ direction, the band
structure near $E_F$ has some contributions from the $p_z$ orbital
and small contributions from the $d_{z^2}$ orbital. They all have
even symmetry and thus can be observed in the $\pi$ experimental
geometry. The small ellipsoidal Fermi surface near zone center is
mainly contributed by the $d_{z^2}$ orbital for FeAs-based
compounds, while for Fe(Te$_{0.66}$Se$_{0.34}$), Te/Se $p_z$ orbital plays
an important role.
To have a more quantitative picture, we have listed the
contributions of various orbitals to the states at $E_F$ in Table. I, which are the coefficients of the calculated corresponding Bloch wavefunctions projected into the orbitals.

%

\section{Discussion}


Although the chalcogen ions contribute little spectral  to the
density-of-states (DOS) near the Fermi energy, the  Fe $3d$ related
band structure in iron-chalcogenides  does show significant
difference compared with that of iron-pnictides. Recently, it is
even proposed that the polarization of the As $p$ orbitals might be
the cause of the unconventional superconductivity in FeAs-based
superconductors \cite{Sawatzky}. Therefore in this regard, the
electronic consequences related to the chalcogen or pnicogen anions
in the iron-based superconductors are particularly interesting to
explore.

In general, Fe$_{1.04}$(Te$_{0.66}$Se$_{0.34}$) has similar Fermi
surface and band structure as the iron-pnictides
\cite{Zhang09B,Zhang09A}. However, there are some important
differences. For example, the three bands near $\Gamma$ are well
separated in this iron-chalcogenides, each with distinct symmetry.
On the other hand in recent polarization dependence studies of the
BaFe$_{1.82}$Co$_{0.18}$As$_{2}$, only two features around $\Gamma$
were observed \cite{Zhang09B}. Moreover, the inner feature is a
mixture of orbitals of both even and odd symmetries, while the outer
feature is even in symmetry. Moreover, our calculations show that
the Te $5p$ orbitals contribute to the density of states near the
Fermi energy, while Fe $3d_{z^2}$ orbital contributes very little in
this iron-chalcogenide. Furthermore, there is a small ellipsoidal
Fermi surface near the zone center of
Fe$_{1.04}$(Te$_{0.66}$Se$_{0.34}$), while for iron-pnictides, such
a small Fermi pocket has not been unambiguously observed in the
paramagnetic normal state.

\begin{figure}[t!]
\includegraphics[width=8.5cm]{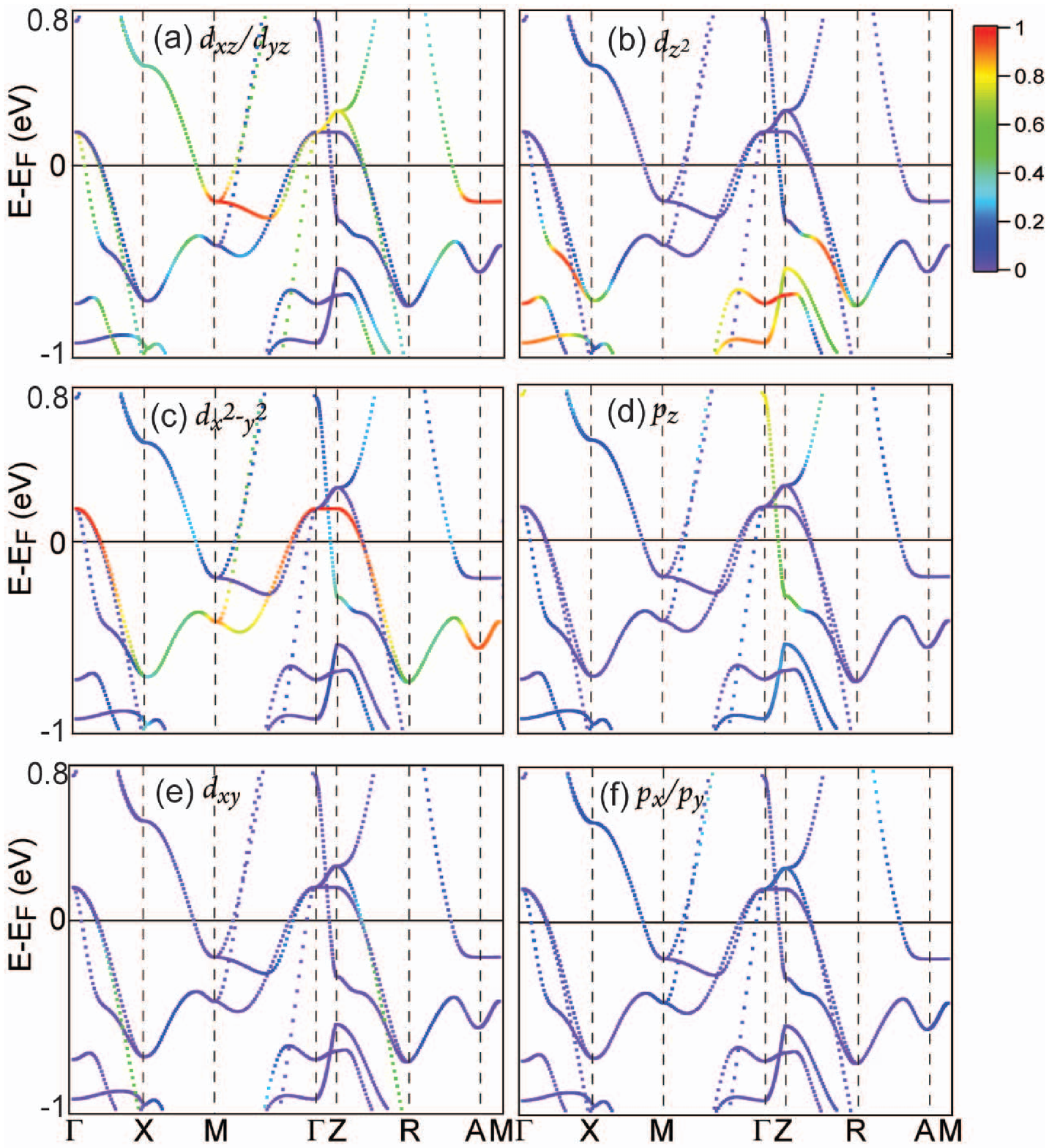}
\caption{(color online) Contributions of various Fe $3d$ and Te/Se
$p$ orbitals to the calculated band structure of
Fe(Te$_{0.66}$Se$_{0.34}$).} \label{LDAorbital}
\end{figure}

Compared with the electronic structure of Fe$_{1+y}$Te obtained
earlier \cite{Hasan}, the Fe$_{1.04}$(Te$_{0.66}$Se$_{0.34}$)
electronic structure behaves differently in the following two
aspects. First, three bands ${\alpha}$, $\beta$, and $\gamma$, are
clearly observed around $\Gamma$ for
Fe$_{1.04}$(Te$_{0.66}$Se$_{0.34}$), whereas only two bands were
distinguished in Fe$_{1+y}$Te. Secondly, a weak Fermi surface was
observed around X-point in Fe$_{1+y}$Te, which was argued to be a
folded Fermi surface by the spin density wave. For
Fe$_{1.04}$(Te$_{0.66}$Se$_{0.34}$), neutron scattering experiments
have found incommensurate short-range magnetic order below 50 K
\cite{Mao}, however, our measurements with two different photon
polarizations confirm the absence of states near $E_F$ around X.
Furthermore, no band splitting that is associated with the SDW in
BaFe$_{2}$As$_{2}$ and SrFe$_{2}$As$_{2}$ is observed here. This
might suggest that such a short range magnetic order should be very
weak in Fe$_{1.04}$(Te$_{0.66}$Se$_{0.34}$).

\begin{table}
\caption{\label{tab:table1} The contributions of Fe $3d$ and Te/Se
$p$ orbitals to the bands in Fe(Te$_{0.66}$Se$_{0.34}$)
near $E_F$ along the $\Gamma-X$ direction. }
\begin{ruledtabular}
\begin{tabular}{lccccccc}

         & $d_{z^2}$   & $d_{xz}$  & $d_{yz}$  &  $d_{x^2-y^2}$ &  $d_{xy}$   &   $p_z$  &
        $p_x+p_y$\\
 $\alpha$  & 0.0391 & 0.6702& 0 & 0.0182 & 0  & 0.0791 & 0.0767 \\
 $\beta$   & 0  & 0& 0.6579 & 0  &  0.2359 & 0 & 0.0463  \\
$\gamma$   & 0.0033 & 0.0434& 0 & 0.9391 & 0  & -0.0001 & 0.0069

\end{tabular}
\end{ruledtabular}
\end{table}


\section{Conclusion}

To summarize, we have  studied the electronic structure of
Fe$_{1.04}$(Te$_{0.66}$Se$_{0.34}$). Both the ARPES and DFT
calculations reveal one inner closed Fermi pocket and two outer
cylindrical Fermi surfaces near $\Gamma$, and two electron-like
Fermi surfaces near the M-point. There are no states near the Fermi
energy around the X-point. Polarization dependence measurements
further elucidate the symmetry of the band structure. The ARPES
results qualitatively agree with the DFT calculations. Compared with
the iron-pnictides, although many aspects of the band structures are
similar, there are also significant differences, particularly in
their electronic structures near $\Gamma$ at the paramagnetic normal
state. Our results provide a comprehensive picture on the electronic
structure of Fe$_{1.04}$(Te$_{0.66}$Se$_{0.34}$), and shed new light
on the role of anions in iron-based superconductors.

\acknowledgements{Part of this work was  performed at the Surface
and Interface Spectroscopy beamline, Swiss Light Source, Paul
Scherrer Institute, Villigen, Switzerland. We thank C. Hess and F. Dubi for technical support. This work was supported by the NSFC, MOE, MOST (National Basic Research Program No.2006CB921300), and STCSM of China.}

\end{document}